# Anomalous Nernst effect on a magnetically doped topological insulator surface: A Green's function approach

Yan-Yan Yang,[1] Ming-Xun Deng,[1,*] Wei Luo,[2] R. Ma,[3] Shi-Han Zheng,[1] and Rui-Qiang Wang[1,†]
[1]*Guangdong Provincial Key Laboratory of Quantum Engineering and Quantum Materials,
ICMP and SPTE, South China Normal University, Guangzhou 510006, China*
[2]*School of Science, Jiangxi University of Science and Technology, Ganzhou 341000, China*
[3]*Jiangsu Key Laboratory for Optoelectronic Detection of Atmosphere and Ocean,
Nanjing University of Information Science and Technology, Nanjing 210044, China*

By generalizing the Kubo-Streda formula for calculating electrical conductivities to the thermoelectric coefficients, we theoretically study the anomalous Nernst effect (ANE) on the surface of a topological insulator induced by a finite concentration of magnetic impurities. The ANE is found to be modulated by the impurity scattering and thermal fluctuations, simultaneously, and so exhibits rich structures in the energy space. While the anomalous Hall conductivity is half-integer quantized with the Fermi level across the magnetic-impurity-induced gap, the anomalous Nernst signal (ANS) is fully suppressed and the thermopower is linear dependent on the Fermi energy. Around the magnetic-impurity-induced localized levels, the ANS and thermopower are resonant enhanced. The suppression and enhancement of the thermoelectric coefficients will compete with each other as the magnetic impurity potential increases continually. More interestingly, when a finite charge potential is included, the resonant peaks of the ANS and thermopower will be renormalized, making the signs of the ANS and thermopower tunable by the strength of the charge potential.

## I. INTRODUCTION

The research of high efficient heat-electricity conversion in low-dimension materials has become an active subject [1–4] and, recently, great interest is triggered in the thermoelectric transports in topological materials [5–9]. Topological insulators (TIs) have a bulk band gap separating the conduction and valence bands, but, as different from ordinary insulators, they support topological protected gapless boundary states inside the bulk band gap [10,11]. The boundary states in three-dimensional (3D) TIs form a two-dimensional (2D) topological surface, on which the states are characterized by the linear Dirac dispersion. The spins of the TI surface's electrons are locked to their momenta in a chiral structure [12–16]. Due to their interesting fundamental physics and potential applications, a wide variety of TIs, e.g., $Bi_xSb_{1-x}$, $Bi_2Se_3$, $Bi_2Te_3$, and $Sb_2Te_3$, have been explored [15–21], among which, $Bi_2Te_3$ and $Bi_2Se_3$ have been demonstrated to be excellent thermoelectric materials with high thermoelectric figure of merit [6,7].

Impurities can substantially influence TI surface's spectrum, so the thermoelectric transport strongly depends not only on the thermal activation, but also on the impurity scattering mechanism. Theoretically, doping magnetic impurities on the TI surface, which breaks time-reversal (TR) symmetry for the surface states, would gap the spectrum of the surface fermions. While Black-Schaffer *et al.* [22] addressed that the energy gap would be filled rapidly by nonmagnetic scattering of the impurities, Deng *et al.* found that, for strong magnetic-impurity potentials, the gap could be filled even by the magnetic scattering itself, and so proposed to study the gap-opening effect of the magnetic impurities by the anomalous Hall effect (AHE) [23]. The magnetic-impurity-modulated AHE in Mn-doped $Bi_2Se_3$ has been observed by recent experiment [24]. Apart from the AHE [25], many other exotic phenomena, such as the anomalous Nernst effect (ANE) [26], are also observable, when the TR symmetry is broken specifically for topological systems. When a temperature gradient $\nabla T$ is applied to these systems, there will exist a measurable electrical current along the longitudinal and transverse directions. Therefore, the charge current density (CCD) $\mathbf{j}^c = (j_x^c, j_y^c)$ can be expressed as $j_\alpha^c = \sum_{\lambda=x,y}[\sigma_{\alpha\lambda}E_\lambda + \alpha_{\alpha\lambda}(-\partial_\lambda T)]$, where $\sigma_{\alpha\lambda}$ and $\alpha_{\alpha\lambda}$ are, respectively, the electrical and thermoelectric conductivity tensors. In the thermoelectric experiment, the thermopower $S_{xx}$ and Nernst signal $S_{yx}$ are measured as the voltage drop induced by the temperature gradient. By tuning the external electrical field, one can obtain the voltage drop when the CCD vanishes, i.e., $\mathbf{j}^c = 0$. As a consequence, the thermoelectric coefficients, defined as $S_{\alpha\gamma} = E_\alpha/\partial_\gamma T$, can be read as

$$S_{\alpha\gamma} = \sum_{\rho=x,y}[\hat{\sigma}^{-1}]_{\alpha\rho}\alpha_{\rho\gamma}. \quad (1)$$

For the case of nonresonant impurity scattering, Berry-curvature-mediated anomalous transports, generated by a mechanical or statistical force [27–29], have been established for 2D systems. In ferromagnets, Xiao *et al.* [27] developed a semiclassical theory of the Berry-phase effect in anomalous transport driven by a temperature gradient, where the Berry-phase correction entered the orbital magnetization rather than the anomalous velocity [28]. Zhang *et al.* [30,31] discussed the anomalous thermoelectric transport in gapped graphene and underdoped cuprate superconductors. It was found the Nernst response in gapped single and bilayer graphene can be used to create a valley-index polarization, and in underdoped cuprate superconductors, the magnitude of the Nernst effect can be sizable even at temperatures much higher than the superconducting transition temperature. In TIs, Choi *et al.* [32] found the Seebeck coefficient, for Mn-doped $Bi_2Se_3$ with tunable type of carrier, increased monotonically with increasing temperature, and enhanced thermoelectric properties of $Bi_2Te_{3-x}Se_x$ were observed by Soni *et al.* [33]. While tremendous efforts have been devoted to investigation of the thermoelectric transports in various systems [30–39], a detail theoretical study of the thermoelectric effect on magnetic impurity-doped TI surfaces, including both the resonant impurity scattering and topological effects, is still lacking. Furthermore, when the extrinsic and intrinsic contributions compete with each other, the thermoelectric effect could be very complicated. On the other hand, when thermal fluctuations are considered, the anomalous Nernst signal could be served as a powerful diagnostic of the topological property of the spectrum structure.

*dengmingxun@scnu.edu.cn
†rqwanggz@163.com

In this paper, by generalizing the Kubo-Streda formula for calculating electrical conductivities, e.g., the AHE, to the thermoelectric coefficients, we study the ANE induced by a finite concentration of magnetic impurities deposited on the surface of a TI. We find that, due to the competition and cooperation effect between the thermal fluctuations and impurity scattering, the ANE on the magnetically doped TI surface exhibits very rich structures in the energy space. When the Fermi level lies in the magnetic-impurity-induced gap, while the anomalous Hall conductivity is half-integer quantized, the anomalous Nernst signal (ANS) is fully suppressed and the thermopower is linear scaled with the Fermi energy. Around the magnetic-impurity-induced localized levels, a resonant enhancement of the ANS and thermopower can be observed. For a relatively weak impurity potential, the suppression and enhancement of the ANS are separated far away from each other in energy. However, as the magnetic impurity potential increases, the resonant peaks of the ANS and thermopower will approach the gap and compete with the topological contributions, making the zero plateau of the ANS and linear Fermi-energy dependence of the thermopower less observable for higher temperatures. The nonmagnetic charged impurities can not induce the anomalous transverse transport. However, in the presence of a finite magnetic potential, the charge potential can destroy the symmetrical distribution of the ANS and thermopower through redistributing the localized levels. More interestingly, the signs of the ANS and thermopower are tunable by the strength of the charge potential.

The paper is organized as follows. In Sec. II, we introduce the model and describe the theory. The thermoelectric coefficients for the magnetically doped TI surface are derived specifically in Sec. III. In Secs. IV and V, we study the magnetic-impurity-modulated ANE and transverse thermal conductivity, respectively, and the last section contains a short summary.

## II. MODEL AND THEORY

The electronic states in the vicinity of the Dirac point on an impurity-free TI surface can be described by a low-energy effective Hamiltonian $H_0 = \sum_{\mathbf{k},\alpha\beta} h_{\alpha\beta}(\mathbf{k}) c_{\mathbf{k}\alpha}^\dagger c_{\mathbf{k}\beta}$ with $h(\mathbf{k}) = \hbar v_F \boldsymbol{\sigma} \cdot \mathbf{k}$, where $v_F$ denotes the Fermi velocity, $\mathbf{k} = (k_x, k_y)$ represents the wave vector, and $\boldsymbol{\sigma} = (\sigma_x, \sigma_y)$ stands for the vector of Pauli matrices for spin. In the presence of impurities, the total Hamiltonian of the system is given by $H = H_0 + H_{imp}$, with

$$H_{imp} = \sum_m \int_S c^\dagger(\mathbf{r}) V(\mathbf{r} - \mathbf{r}_m) c(\mathbf{r}) d\mathbf{r}, \quad (2)$$

where $V(\mathbf{r} - \mathbf{r}_m) = (U_0 \sigma_0 + U_M \mathbf{s} \cdot \boldsymbol{\sigma}) \delta(\mathbf{r} - \mathbf{r}_m)$ is the scattering potential of a single impurity located at position $\mathbf{r}_m$, $c^\dagger(\mathbf{r}) = (c_\uparrow^\dagger(\mathbf{r}), c_\downarrow^\dagger(\mathbf{r}))$ is the electron creation operator with $c_\sigma^\dagger(\mathbf{r}) = \frac{1}{\sqrt{N}} \sum_{\mathbf{k}} c_{\mathbf{k}\sigma}^\dagger e^{i\mathbf{k}\cdot\mathbf{r}}$, and $\mathbf{s}$ is the unit vector of the local impurity moment. The magnetic impurities under consideration here are randomly distributed on the TI surface, including both a short-range charge ($U_0$) and magnetic ($U_M \mathbf{s}$) potential. The integral in $H_{imp}$ is performed over the sample, and the summation runs over the possible positions which the impurities may locate in. Since the quantum nature of the impurity spins is not crucial, we would treat the impurities to be classical type, in which the exchange coupling integral $U_{M,0}$ in $V = U_0 \sigma_0 + U_M \mathbf{s} \cdot \boldsymbol{\sigma}$ is kept to be a constant. It is well known that the skew-scattering and side-jump mechanisms are very important to the anomalous transport for correlated and spin-dependent scatters [40,41]. Here, we consider the TI surface is doped with dilute impurities, i.e., the impurities are short range and randomly distributed, such that the correlation between the impurities is negligible, i.e., $\langle V(\mathbf{r}) V(\mathbf{r}') \rangle = 0$. In this case, the main contributions originate from the gap-opening and localization effects of the impurities. In fact, the topological term due to disorder, as discussed in Ref. [40], will always overwhelm the terms due to the band structure, skew scattering, and side jump. Therefore, we can omit the skew-scattering and side-jump mechanisms for simplicity, and this omission would not change the qualitative results.

To maintain the gauge invariance, which is related to charge conservation [42], the group velocity operators are determined by $\hat{v}_{x,y} = -\hbar^{-1} \partial G_\mathbf{k}^{-1}(i\omega_n) / \partial k_{x,y}$, where

$$G_\mathbf{k}(i\omega_n) = \frac{1}{i\hbar\omega_n - h(\mathbf{k}) - \Sigma_\mathbf{k}(i\omega_n)} \quad (3)$$

is the Matsubara Green's function and $\Sigma_\mathbf{k}(i\omega_n)$ is the self-energy introduced by the impurity scattering, with $\omega_n = (2n+1)\pi/\beta$ ($n \in \mathbb{Z}$) being the Matsubara frequencies. By inspecting the poles of the Matsubara Green's function, one can obtain an effective Hamiltonian $h_{eff} = h(\mathbf{k}) + \Sigma_\mathbf{k}(i\omega_n)$ for the system. For the TI surface doped with dilute magnetic impurities, we would adopt a $T$-matrix approach to calculate the self-energy and follow the standard momentum or Brillouin-zone cutoff to regularize the linear Dirac spectrum. The $T$-matrix approach takes into account the multiple scattering events of electrons off single impurity but ignores the electron scattering between impurities, which is justified for lightly doped case where interactions among impurities can be ignored. Although the Ward identities may be violated slightly due to the Brillouin-zone cutoff, as we will show, our results can be accurate to first order in impurity density. Following the regularization scheme [42] developed by Fujimoto *et al.*, one can maintain the Ward identities. However, as shown by Eqs. (8) and (15) of Ref. [42], the vertex function just renormalizes the Fermi velocity by a small quantity in first order of the impurity density. Therefore, the vertex corrections to the anomalous transport coefficients will be higher-order effects (at least second order in impurity density), which is negligible for low-concentration doping case. Consequently, introducing the vertex corrections here or not has little effect on the conclusions and we would omit contributions from ladder diagrams in the specific calculation. Accordingly, we can make the approximation [43] $\Sigma_\mathbf{k}(i\omega_n) = \rho_{\text{imp}} T(i\omega_n)$, where

$$T(i\omega_n) = \left[\sigma_0 - V \sum_\mathbf{k} G_0(\mathbf{k}, i\omega_n)\right]^{-1} V \quad (4)$$

represents the $T$ matrix and $\rho_{\text{imp}}$ denotes the impurity density, with $G_0(\mathbf{k}, i\omega_n)$ the impurity-free Green's function for the TI surface. The CCD and energy current density (ECD) operators, accordingly, are given by [43] $\hat{j}^c_{x,y} = e\hat{v}_{x,y}$ and $\hat{j}^E_{x,y} = \{\hat{v}_{x,y}, h_{\text{eff}}\}/2$, respectively. The CCD and ECD tensors can be incorporated into a vector equation, reading as

$$\begin{pmatrix} \mathbf{j}^c \\ \mathbf{j}^E \end{pmatrix} = \begin{pmatrix} \mathbf{L}^{11} & \mathbf{L}^{12} \\ \mathbf{L}^{21} & \mathbf{L}^{22} \end{pmatrix} \begin{pmatrix} \mathbf{E} - \frac{T}{e}\nabla\left(\frac{\mu}{T}\right) \\ T\nabla\left(\frac{1}{T}\right) \end{pmatrix}, \quad (5)$$

where the elements of the transport coefficient matrix are given by the current-current correlation function [43,44]

$$L^{nm}_{\alpha\gamma} = \lim_{s\to 0} \frac{1}{\Omega} \int_0^\infty dt \, e^{-st} \int_0^\beta d\beta' \langle \hat{j}^m_\gamma(-t - i\hbar\beta') \hat{j}^n_\alpha \rangle \quad (6)$$

with $\Omega$ as area of the TI surface. The superscript in $\hat{j}^{m(n)}_{\gamma(\alpha)}$ is used to distinguish the CCD and ECD operators, i.e., $m, n = 1 (= 2)$ corresponds to the CCD (ECD) operator. Consequently, the CCD tensor can be written as $\mathbf{j}^c = \hat{\sigma}(\mathbf{E} - \nabla\mu/e) + \hat{\alpha}(-\nabla T)$, where the electrical and thermoelectric conductivity tensors are given by

$$\sigma_{\alpha\gamma} = L^{11}_{\alpha\gamma}, \quad (7)$$

$$\alpha_{\alpha\gamma} = \frac{1}{T}\left(L^{12}_{\alpha\gamma} - \frac{\mu}{e}L^{11}_{\alpha\gamma}\right). \quad (8)$$

By setting $\mathbf{j}^c = 0$, we solve for the effective electrical field and obtain

$$\mathbf{E} - \nabla\mu/e = \hat{\sigma}^{-1}\hat{\alpha}\nabla T, \quad (9)$$

from which we can find the thermoelectric coefficient tensor [35,36,45]

$$S_{\alpha\gamma} = (E_\alpha - \partial_\alpha\mu/e)/\partial_\gamma T = \sum_{\rho=x,y}[\hat{\sigma}^{-1}]_{\alpha\rho}\alpha_{\rho\gamma}. \quad (10)$$

Subsequently, we can derive the thermopower $S_{xx}$ and Nernst signal $S_{yx}$ to be

$$S_{xx} = \frac{\alpha_{xx}}{\sigma_{xx} + \sigma_{yx}\tan\theta^y_H} + \frac{\alpha_{yx}}{\sigma_{yx} + \sigma_{xx}\cot\theta^y_H}, \quad (11)$$

$$S_{yx} = \frac{\alpha_{yx}}{\sigma_{yy} + \sigma_{yx}\tan\theta^x_H} - \frac{\alpha_{xx}}{\sigma_{yx} + \sigma_{yy}\cot\theta^x_H}, \quad (12)$$

where we have used the symmetry $\sigma_{xy} = -\sigma_{yx}$ and the definitions $\theta^{x,y}_H = \tan^{-1}(\sigma_{yx}/\sigma_{xx,yy})$ of the Hall angles. For isotropy systems, $\sigma_{yy} = \sigma_{xx}$ and $\theta^x_H = \theta^y_H = \theta_H$, such that $S_{xx}$ and $S_{yx}$ return to the generalized Mott formula shown in Ref. [46]. If $\theta_H \to 0$, the first term of Eqs. (11) and (12) completely defines the thermopower and Nernst signal, and they recover the formula given in Refs. [30,37] for small Hall angle.

In the following, we will derive the transport coefficients $L^{mn}_{\alpha\gamma}$ specifically. For undoped systems $\Sigma_\mathbf{k}(i\omega_n) \to 0$, the eigenvalue problem is exactly solvable, such that $L^{mn}_{\alpha\gamma}$ just takes the form of the Kubo formula, i.e.,

$$L^{nm}_{\alpha\gamma} = \frac{i\hbar}{\Omega}\sum_{\mathbf{k},\mu\nu} \frac{\hat{j}^n_{\alpha,\mu\nu}\hat{j}^m_{\gamma,\nu\mu}(f_{k,\mu} - f_{k,\nu})}{(E_{k,\mu} - E_{k,\nu})(E_{k,\mu} - E_{k,\nu} + i0^+)}, \quad (13)$$

where $f_{k,\nu} = (1 + e^{\beta(E_{k,\nu} - E_F)})^{-1}$ is the Fermi-Dirac distribution function and $\hat{j}^n_{\alpha,\mu\nu} = \langle \mathbf{k}, \mu | \hat{j}^n_\alpha | \mathbf{k}, \nu \rangle$, with $E_{k,\nu}$ and $|\mathbf{k}, \nu\rangle$ being the eigenenergy and wave function for the system. As it shows, the transverse transport coefficients are Berry-curvature mediated, which requires the system to be TR-symmetry broken. Consequently, for the present system, the transverse transport coefficients are expected to vanish if $U_M = 0$, due to the TR symmetry. In the presence of impurities, an energy-dependent self-energy would be introduced, which makes it difficult to determine the dispersion analytically, especially around the localized states. Therefore, the expressions for the transport coefficients, turning to be very complicated, are beyond the usual Kubo formula. In this situation, it is constructive to derive Eq. (6) in terms of the Matsubara Green's function. According to the definition $G_{\mathbf{k}\rho,\mathbf{k}'\mu}(\tau - \tau') = -\langle T_\tau c_{\mathbf{k}\rho}(\tau) c^\dagger_{\mathbf{k}'\mu}(\tau')\rangle$, we can express Eq. (6) as

$$L^{nm}_{\alpha\gamma} = -\frac{1}{\Omega}\lim_{\omega \to 0^+}\frac{1}{\beta}\sum_{\mathbf{k},\upsilon}\text{Tr}\big[\hat{j}^m_\gamma \partial_\omega G_\mathbf{k}(i\omega_\upsilon + i\omega)\hat{j}^n_\alpha$$
$$\times G_\mathbf{k}(i\omega_\upsilon) + \hat{j}^m_\gamma G_\mathbf{k}(i\omega_\upsilon)\hat{j}^n_\alpha \partial_\omega G_\mathbf{k}(i\omega_\upsilon - i\omega)\big], \quad (14)$$

where $G_\mathbf{k}(i\omega_\upsilon)$ is the Fourier transformation of $G_\mathbf{k}(\tau - \tau')$. For further calculations, it is favorable to express the Matsubara frequency summation by real-frequency integrals. To this end, let us consider contour integral $I = \frac{1}{2\pi i}\lim_{R\to\infty}\oint_R dz\,\Phi(z)$ with $\Phi(z) = f(z)\text{Tr}[\hat{j}^m_\gamma \partial_\omega G_\mathbf{k}(z + i\omega)\hat{j}^n_\alpha G_\mathbf{k}(z)]$, as depicted in Fig. 1. The contour integral, within the framework of the residue theorem, is easy to obtain as $I = \sum_\upsilon \text{Res}[\Phi(z), i\omega_\upsilon] + \sum_i \text{Res}[\Phi(z), \omega_i]$, where $i\omega_\upsilon$ are singularities of $\Phi(z)$ in the imaginary axis due to $f(z)$, and $\omega_i$ are singularities along the real axis contributed from the Matsubara Green's function. For $\Sigma_\mathbf{k}(i\omega_n) \to 0$, the singularities of the Matsubara Green's function are well defined, and Eq. (14) recovers Eq. (13). In general situations, where the self-energy is finite and energy dependent, we can use the residue theorem again, to express $\sum_i \text{Res}[\Phi(z), \omega_i]$ by a contour integral along the real axis. To perform this procedure, we keep $\omega$ finite during the analytical continuation and take the limit $\omega \to 0$ after the analytical continuation has been performed. Therefore, for $z = \epsilon + i0^+$ (the lines above the real axis), both $G_\mathbf{k}(\epsilon + i0^+ + i\omega)$ and $G_\mathbf{k}(\epsilon + i0^+)$ are retarded Green's functions. For $z = \epsilon - i0^+$ (the lines below the real axis), while $G_\mathbf{k}(\epsilon - i0^+)$ becomes an advance Green's function, $G_\mathbf{k}(\epsilon - i0^+ + i\omega)$ is still a retarded Green's function, since $\omega > 0^+$. Similar procedure can be performed on the second term of Eq. (14). Finally, we can obtain for

$$L_{\alpha\gamma}^{nm} = -\frac{\hbar}{2\pi\Omega}\sum_\mathbf{k}\int_{-\infty}^\infty d\epsilon\, f(\epsilon)\text{Tr}\{\hat{j}_\gamma^m \partial_\epsilon G_\mathbf{k}^R(\epsilon)\hat{j}_\alpha^n[G_\mathbf{k}^R(\epsilon) - G_\mathbf{k}^A(\epsilon)] - \hat{j}_\gamma^m[G_\mathbf{k}^R(\epsilon) - G_\mathbf{k}^A(\epsilon)]\hat{j}_\alpha^n \partial_\epsilon G_\mathbf{k}^A(\epsilon)\}. \quad (15)$$

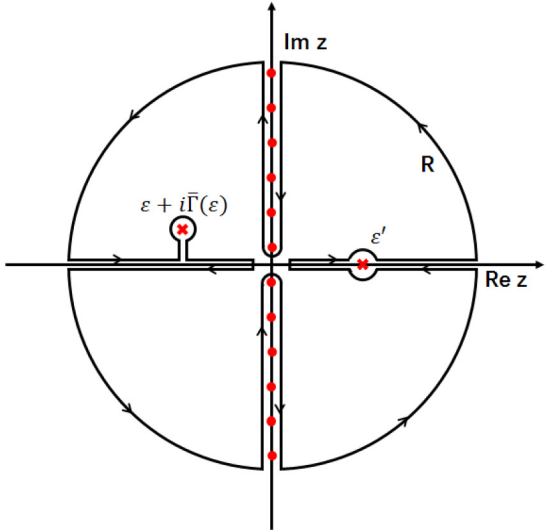

FIG. 1. The contour of $I = \frac{1}{2\pi i}\lim_{R\to\infty}\oint dz\, \Phi(z)$ with $\overline{\Gamma}(\varepsilon) = \text{Im}[\Sigma_\mathbf{k}(\varepsilon + i\omega)] - \omega$, where the red filled circles represent the singularities of $\Phi(z)$ due to the Fermi-Dirac distribution function and the red crosses denote the singularities contributed from Green's functions $G_\mathbf{k}(z + i\omega)$ and $G_\mathbf{k}(z)$. The singularities of $G_\mathbf{k}(z)$, in fact, should be determined by the equation $z = \varepsilon + \Sigma_\mathbf{k}(z)$ self-consistently. For the complicated $z$-dependent self-energy, it is difficult to determine the locations of the singularities of the Green's functions, definitely. We represent the singularities of $\Phi(z)$ by the red crosses and filled circles just for the purpose of illustration.

Here, $G_\mathbf{k}^{R,A}(\epsilon)$ is the retarded and advance Green's functions. If we keep one-half of Eq. (15) and make an integration by parts on the second half, the transport coefficients can be rewritten as $L_{\alpha\gamma}^{nm} = L_{\alpha\gamma}^{nm(\text{I})} + L_{\alpha\gamma}^{nm(\text{II})}$, where

$$L_{\alpha\gamma}^{nm(\text{I})} = \frac{\hbar}{4\pi\Omega}\sum_\mathbf{k}\int_{-\infty}^\infty d\epsilon\, \partial_\epsilon f(\epsilon)\text{Tr}\{\hat{j}_\gamma^m[G_\mathbf{k}^A(\epsilon) - G_\mathbf{k}^R(\epsilon)]$$
$$\times \hat{j}_\alpha^n G_\mathbf{k}^A(\epsilon) - \hat{j}_\gamma^m G_\mathbf{k}^R(\epsilon)\hat{j}_\alpha^n[G_\mathbf{k}^A(\epsilon) - G_\mathbf{k}^R(\epsilon)]\} \quad (16)$$

and

$$L_{\alpha\gamma}^{nm(\text{II})} = \frac{\hbar}{4\pi\Omega}\sum_\mathbf{k}\int_{-\infty}^\infty d\epsilon\, f(\epsilon)\text{Tr}\{\hat{j}_\gamma^m \partial_\epsilon G_\mathbf{k}^A(\epsilon)$$
$$\times \hat{j}_\alpha^n G_\mathbf{k}^A(\epsilon) - \hat{j}_\gamma^m G_\mathbf{k}^A(\epsilon)\hat{j}_\alpha^n \partial_\epsilon G_\mathbf{k}^A(\epsilon) + \hat{j}_\gamma^m G_\mathbf{k}^R(\epsilon)$$
$$\times \hat{j}_\alpha^n \partial_\epsilon G_\mathbf{k}^R(\epsilon) - \hat{j}_\gamma^m \partial_\epsilon G_\mathbf{k}^R(\epsilon)\hat{j}_\alpha^n G_\mathbf{k}^R(\epsilon)\}. \quad (17)$$

A similar formula, i.e., the Kubo-Streda formula, was first shown by Crepieux and Bruno [47], where the conductivity was divided into the contributions of electrons at the Fermi level $\sigma_{\alpha\gamma}^{(\text{I})} = L_{\alpha\gamma}^{11(\text{I})}$ and filled states below the Fermi level $\sigma_{\alpha\gamma}^{(\text{II})} = L_{\alpha\gamma}^{11(\text{II})}$. The Kubo-Streda formula is widely-used for studies of the AHE [23,48–51].

### III. THERMOELECTRIC TRANSPORT COEFFICIENTS FOR THE MAGNETICALLY DOPED TI SURFACE

For dilute magnetically doped TI surfaces, the $T$ matrix can be derived as

$$T(i\omega_n) = \zeta_+(i\omega_n)\sigma_0 + \zeta_-(i\omega_n)\mathbf{s}\cdot\boldsymbol{\sigma}, \quad (18)$$

where

$$\zeta_\pm(i\omega_n) = \frac{1}{2}\left[\frac{1}{1/U_+ - g(i\omega_n)} \pm \frac{1}{1/U_- - g(i\omega_n)}\right] \quad (19)$$

with $U_\pm = U_0 \pm U_M$, $g(i\omega_n) = \frac{i\omega_n N}{4\pi(\hbar v_F)^2}\ln\frac{(i\omega_n)^2}{(i\omega_n)^2-\Lambda^2}$, and $\Lambda$ a high-energy cutoff. Accordingly, we can obtain for the Matsubara Green's function

$$G_\mathbf{k}(i\omega_n) = \frac{1}{2}\sum_{\eta=\pm}\frac{\sigma_0 + P(i\omega_n)/\xi_\eta(i\omega_n)}{i\widetilde{\omega}_n - \xi_\eta(i\omega_n)}, \quad (20)$$

where $\xi_\eta(i\omega_n) = \eta\sqrt{(\hbar v_F)^2(\widetilde{k}_x^2 + \widetilde{k}_y^2) + \Delta^2(i\omega_n)}$ and

$$P(i\omega_n) = \begin{pmatrix} \Delta(i\omega_n) & \hbar v_F(\widetilde{k}_x - i\widetilde{k}_y) \\ \hbar v_F(\widetilde{k}_x + i\widetilde{k}_y) & -\Delta(i\omega_n) \end{pmatrix} \quad (21)$$

with

$$i\widetilde{\omega}_n = i\omega_n - \rho_{\text{imp}}\zeta_+(i\omega_n), \quad (22)$$

$$\widetilde{k}_{x,y} = k_{x,y} + \rho_{\text{imp}}\frac{\zeta_-(i\omega_n)}{\hbar v_F}s_{x,y}, \quad (23)$$

and

$$\Delta(i\omega_n) = \rho_{\text{imp}}\zeta_-(i\omega_n)s_z. \quad (24)$$

The retarded and advance Green's functions can be obtained by the analytic continuation $G_\mathbf{k}^{R,A}(\epsilon) = G_\mathbf{k}(i\omega_n)|_{i\omega_n \to \epsilon^\pm}$, with $\epsilon^\pm = \epsilon \pm i0^+$. As shown by Eq. (23), the averaged effect of the in-plane impurity moment here just causes a translation of $k_{x,y}$, so that the shape of the Fermi surface remains unchanged and, due to the translational symmetry, it will vanish for Eqs. (16) and (17) after the momentum summation. On the other hand, the topological nontrivial contribution to the anomalous transport is attributed to the Berry curvature, which is nonvanishing for a gapped TI surface, due to the requirement of broken TR symmetry. By inspecting the poles of the retarded Green's function, we find no gap when the averaged impurity moment is confined in the $x$-$y$ plane, so that we would consider a favorable situation, where the averaged impurity moment is aligned in the normal direction to the $x$-$y$ plane. Subsequently, to first order in $\rho_{\text{imp}}$, we can arrive at

$$L_{yx}^{nm(\text{I})} = \frac{e^2}{h}\int_{-\infty}^{\infty} d\epsilon [-\partial_\epsilon f(\epsilon)] \frac{\chi_{mn}}{\pi} \frac{(\pi-\phi)\,\text{Im}[\tilde{\epsilon}^-\Delta(\epsilon^+)]}{|\mathcal{F}(\epsilon^+)|\sin\phi} \quad (25)$$

and

$$L_{yx}^{nm(\text{II})} = \frac{e^2}{h}\int_{-\infty}^{\infty} d\epsilon\, f(\epsilon) \frac{\chi_{mn}}{\pi}\,\text{Im}\,\frac{\tilde{\epsilon}^+\partial_\epsilon\Delta(\epsilon^+) - \Delta(\epsilon^+)\partial_\epsilon\tilde{\epsilon}^+}{\mathcal{F}(\epsilon^+)}, \quad (26)$$

where $\mathcal{F}(\epsilon^+) = (\tilde{\epsilon}^+)^2 - \Delta^2(\epsilon^+)$ with argument $\phi$, $\tilde{\epsilon}^\pm = \epsilon^\pm - \rho_{\text{imp}}\zeta_+(\epsilon^\pm)$, and

$$\chi = \begin{pmatrix} 1 & \epsilon/e \\ \epsilon/e & \epsilon^2/e^2 \end{pmatrix}. \quad (27)$$

From Eq. (19), it is easy to find

$$\partial_\epsilon[\zeta_+(\epsilon^+) \pm \zeta_-(\epsilon^+)] = [\zeta_+(\epsilon^+) \pm \zeta_-(\epsilon^+)]^2 \partial_\epsilon g(\epsilon^+), \quad (28)$$

so that we can further reduce $L_{yx}^{nm(\text{II})}$ to be

$$L_{yx}^{nm(\text{II})} = -\frac{e^2}{h}\int_{-\infty}^{\infty} d\epsilon\, f(\epsilon)\frac{\chi_{mn}}{\pi}$$
$$\times \text{Im}\left[\frac{1 - 2\epsilon_R\zeta_+(\epsilon^+)\partial_\epsilon g(\epsilon^+)}{\mathcal{F}(\epsilon^+)}\Delta(\epsilon^+)\right], \quad (29)$$

where

$$\epsilon_R = \epsilon^+ - \rho_{\text{imp}}\frac{\zeta_+^2(\epsilon^+) - \zeta_-^2(\epsilon^+)}{2\zeta_+(\epsilon^+)}. \quad (30)$$

The longitudinal transport coefficients are obtained as

$$L_{xx}^{nm} = \frac{1}{h}\int_{-\infty}^{\infty} d\epsilon[-\partial_\epsilon f(\epsilon)]\frac{\chi_{mn}}{2\pi}$$
$$\times \left[1 + (\pi-\phi)\frac{|\tilde{\epsilon}^+|^2 - |\Delta(\epsilon^+)|^2}{|\mathcal{F}(\epsilon^+)|\sin\phi}\right]. \quad (31)$$

## IV. MAGNETIC-IMPURITY-MODULATED ANE

The magnetically doped TI surface can host an interesting ANE, which will be studied in this section. The Nernst effect has been extensively explored in a variety of systems such as graphene [30,36], off-resonant-light-driving TI surface [52], Weyl semimetals [46], and phosphorene [45]. As seen from Eqs. (25) and (29), the anomalous transverse transport is directly proportional to the effective Zeeman-type field $\Delta(\epsilon^+) = \rho_{\text{imp}}\zeta_-(\epsilon^+)s_z$. The effective Zeeman-type field is produced by the collective effect of a finite concentration of magnetic impurities, determined by the impurity density $\rho_{\text{imp}}$ and magnetic impurity potential $U_M s_z$.

For relatively weak impurity potential, i.e., $1/U_\pm \gg \text{Re}\,g(\epsilon^+)$, the localized states induced by the impurity scattering are far away from the Dirac point in energy, such that we can safely make the approximations $\zeta_+(\epsilon) \approx U_0$ and $\zeta_-(\epsilon) \approx U_M$. Therefore, we can receive an effective Hamiltonian $h_{\text{eff}} = h(\mathbf{k}) + \rho_{\text{imp}}(U_0\sigma_0 + U_M\mathbf{s}\cdot\boldsymbol{\sigma})$. As can be seen, the effect of a finite concentration of short-range magnetic impurities, in the weak impurity scattering regime, is similar to that of a ferromagnet capping on the TI surface. Consequently, the magnetic impurities would gap the spectrum of the surface

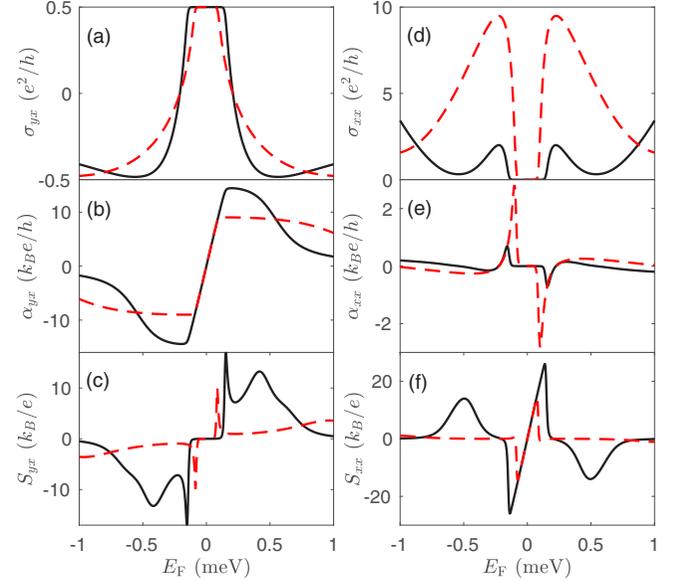

FIG. 2. The transverse (left panel) and longitudinal (right panel) transport coefficients as functions of the Fermi energy, for $U_M = -1.0$ (red dashed lines) and $-2.0$ (dark solid lines), with other parameters set as $U_0 = 0$, $k_B T = 5\times 10^{-3}$ ($T \sim 0.06$ K), $\rho_{\text{imp}} = 0.1$, and $\Lambda = 300$ ($\sim 300$ meV corresponding to the bulk band gap [16]). The unit of the averaged impurity potential is meV.

electrons, and thus result in interesting anomalous transport phenomena to the TI surface.

At low temperatures, we can further reduce

$$L_{yx}^{nm(\text{I})} = \frac{e^2}{h}\frac{\rho_{\text{imp}}U_M s_z}{4}\chi_{mn}|_{\epsilon\to E_F}$$
$$\times \frac{\theta(E_F - \rho_{\text{imp}}U_+) - \theta(\rho_{\text{imp}}U_- - E_F)}{E_F - \rho_{\text{imp}}U_0} \quad (32)$$

and

$$L_{yx}^{nm(\text{II})} = \frac{e^2}{h}\frac{f(\rho_{\text{imp}}U_+)\chi_{mn}^+ - f(\rho_{\text{imp}}U_-)\chi_{mn}^-}{2}, \quad (33)$$

where $\chi_{mn}^\pm = \chi_{mn}|_{\epsilon\to\rho_{\text{imp}}U_\pm}$ and $\theta(x)$ is the Heaviside function. As shown by Eqs. (32) and (33), while the anomalous Hall conductivity $\sigma_{yx} = L_{yx}^{11(\text{I})} + L_{yx}^{11(\text{II})}$ is half-integer quantized ($e^2/2h$) with the Fermi level across the impurity-induced gap, the transverse thermoelectric conductivity $\alpha_{yx} = (E_F - \mu)\sigma_{yx}/eT$ is linear scaled with the Fermi level, which can be easily observed in Figs. 2(a) and 2(b). On the other hand, due to the vanishing density of states (DOS), the longitudinal conductivity $\sigma_{xx} = L_{xx}^{11}$ disappears in the gap. Consequently, $\theta_H \to \pi/2$ for the insulated regime. In this limit, the second term of Eqs. (11) and (12) dominates the Nernst signal $S_{yx} = -\alpha_{xx}/\sigma_{yx}$ and the thermopower $S_{xx} = \alpha_{yx}/\sigma_{yx}$ with $\alpha_{xx} = \alpha_{yx}|_{y\to x}$. Accordingly, the Nernst signal is fully suppressed and the thermopower is linear scaled with $E_F$, when the Fermi level lies in the gap, as displayed in Figs. 2(c) and 2(f), respectively.

In the metallic regime, where the Fermi level is out of the energy gap, the properties of the transport coefficients are dominated by the contribution of the electrons at the Fermi surface. With the Fermi level keeping away from the gap, the DOS at the Fermi surface increases, and $\sigma_{xx}$ will increase linearly with $|E_F|$ for an energy-independent scattering rate, e.g., $\Sigma(\epsilon) = i\Gamma$. If $E_F \gg 0$, $\sigma_{xx} \gg \sigma_{yx} \to 0$ and then $\theta_H \to 0$. In this case, the first term of Eqs. (11) and (12) will dominate the Nernst signal and thermopower. With increasing the impurity potential, the gap becomes larger and, simultaneously, the impurity-induced localized states would approach the Dirac point, leading to filling of the energy gap. When $1/U_\pm - \text{Re} \, g(E_F^+) = 0$, the DOS will develop a resonant peak at the Fermi level. While the DOSs are resonant enhanced, the longitudinal conductivity is heavily suppressed around the resonant levels. In fact, the electronic states around the resonant levels are virtual bound states. The stronger the impurity potential is, the more severe the localization effect becomes. As a result, the longitudinal conductivity, as shown in Fig. 2(d), is very sensitive to the impurity. However, the extrinsic anomalous Hall conductivity in the vicinity of the localized states, with sign opposite to the intrinsic contribution, increases with the magnetic impurity potential. Therefore, around the localized levels, the Hall angle $\theta_H$ will increase with the magnetic impurity potential, and the contributions of both terms of Eqs. (11) and (12) are additive. As indicated in Figs. 2(c) and 2(f), the Nernst signal and thermopower show a resonant enhancement around the localized levels. At the edges of the gap, the Nernst signal exhibits van Hove singularities. As the magnetic impurity potential increases, the resonant peaks approach the gap, and the van Hove singularities are enhanced, gradually.

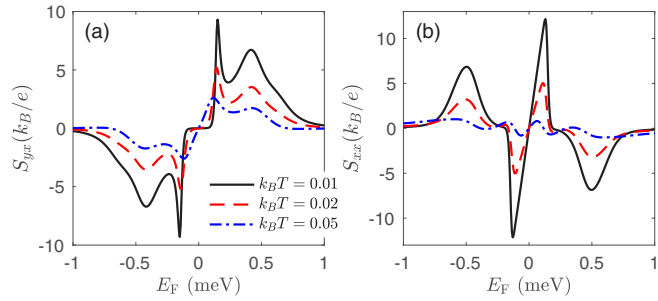

FIG. 3. (a) The Nernst signal and (b) thermopower versus the Fermi energy, with $U_M = -2.0$ and $k_B T = 0.01, 0.02, 0.05$ for the blue dashed-dotted, red dashed, and solid dark curves, respectively. Other parameters are the same as Fig. 2.

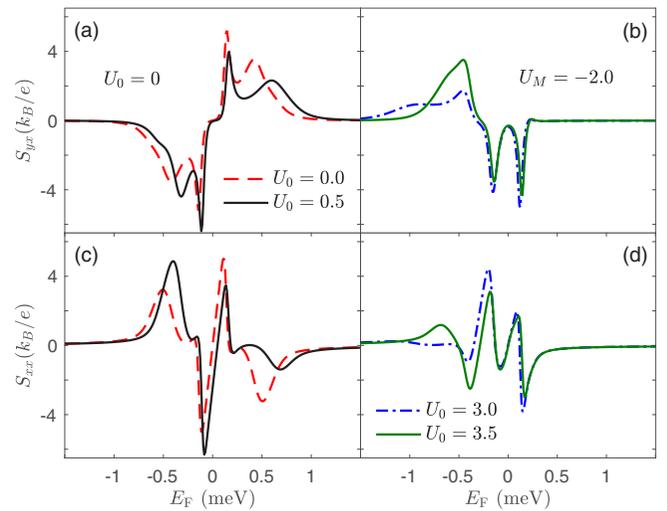

FIG. 4. The Nernst signal (upper panel) and thermopower (lower panel) as functions of the Fermi energy for varied $U_0$, with fixed magnetic impurity potential $U_M = -2.0$ and temperature $k_B T = 0.02$. Other parameters are the same as Fig. 2.

For higher temperatures, the impurity-induced gap can be filled by the thermal excitations. Consequently, as the temperature increases, the suppression of the longitudinal conductivity in the gap will be lifted, accompanied with the anomalous Hall conductivity plateau turning to a peak and shifting away from the half-integer quantized values. As a result, with the increment on the temperature, the zero plateau of $S_{yx}$ would be filled by the thermal excitations, and the linearly scaled character of $S_{xx}$ becomes less obvious. Since the localized electrons can be excited by the thermal fluctuations, $\sigma_{xx}$ will increase with the temperature and, as a result, the higher the temperature is, the lower the resonant peaks of $S_{yx}$ and $S_{xx}$ become, as can be seen from Fig. 3.

The charge impurities, as demonstrated in $h_{\text{eff}}$, simply renormalize the energy position of the Dirac point for weak impurity scattering. Accordingly, nonmagnetic charged impurities alone cannot induce the anomalous transverse transport. This can be inferred from Eqs. (25) and (29), directly. For $U_M = 0$ and $\Delta(\epsilon) = 0$, both $L_{yx}^{mn(\text{I})}$ and $L_{yx}^{mn(\text{II})}$ vanish. In the presence of a finite magnetic impurity potential, the charge potential can destroy the symmetrical distribution of the Nernst signal and thermopower through redistributing the localized levels, as shown in Fig. 4. With increasing $U_0$, the localized level for $E_F < 0$ will approach the Dirac point, accompanied with an increasing and narrowing of the corresponding resonant peaks of $S_{yx}$ and $S_{xx}$. The localized level for $E_F > 0$ behaves in an opposite way. As $U_0$ increases, it shifts away from the Dirac point, and the resonant peaks of $S_{yx}$ and $S_{xx}$ vanish gradually. As a result, the original symmetric distribution of the Nernst signal and thermopower with respect to the Dirac point are destroyed. The localized level for $E_F > 0$ will entirely disappear if $|U_0| \geq |U_M|$, i.e., $U_\pm > 0$, followed by the emergence of another localized level on the $E_F < 0$ side, as shown by the right panel of Fig. 4. If $U_0$ changes sign, the behavior of the resonant peaks for $E_F < 0$ and $E_F > 0$ will interchange. More interestingly, from Figs. 4(b) and 4(d), we see that the sign of the Nernst signal and thermopower is tunable by the strength of the charge potential.

As discussed above, charge impurity potential alone can not induce the anomalous transport, but, in the presence of a finite magnetic moment, it modifies the signs of the transport coefficients through redistributing the energy positions of the localized states. Introducing correlations does not change the energy locations of the localized states and, therefore, we can expect that introducing correlation effects to the present system would not affect the signs of the transport coefficients.

## V. MAGNETIC-IMPURITY-MODULATED TRANSVERSE THERMAL CONDUCTIVITY

In terms of $\mathbf{j}^c$ and $\mathbf{j}^E$, the heat current density (HCD) is defined as [43] $\mathbf{j}^Q = \mathbf{j}^E - \mu \mathbf{j}^c / e$ and the thermal conductivity tensor $\widehat{\kappa}$ can be determined from the HCD by $j_{\alpha\gamma}^Q = \kappa_{\alpha\gamma}(-\partial_\gamma T)$. Usually, the thermal conductivity includes both electron and photon contributions. Here, we mainly concern on the electron contribution, modulated by the magnetic impurity scattering effect. According to Eq. (5), it is easy to find

$$\begin{pmatrix} \mathbf{j}^c \\ \mathbf{j}^Q \end{pmatrix} = \begin{pmatrix} \widehat{\sigma} & \widehat{\alpha} \\ \widehat{\overline{\alpha}} & \widehat{\overline{\kappa}} \end{pmatrix} \begin{pmatrix} \mathbf{E} - \nabla\mu/e \\ -\nabla T \end{pmatrix}, \quad (34)$$

where $\widehat{\overline{\alpha}}$ is related to $\widehat{\alpha}$ by Onsager's relation $\widehat{\overline{\alpha}} = T\widehat{\alpha}$ and the matrix elements of $\widehat{\overline{\kappa}}$ can be obtained from

$$\overline{\kappa}_{\alpha\gamma} = \frac{1}{T}\left(L_{\alpha\gamma}^{22} - 2\frac{\mu}{e}L_{\alpha\gamma}^{12} + \frac{\mu^2}{e^2}L_{\alpha\gamma}^{11}\right). \quad (35)$$

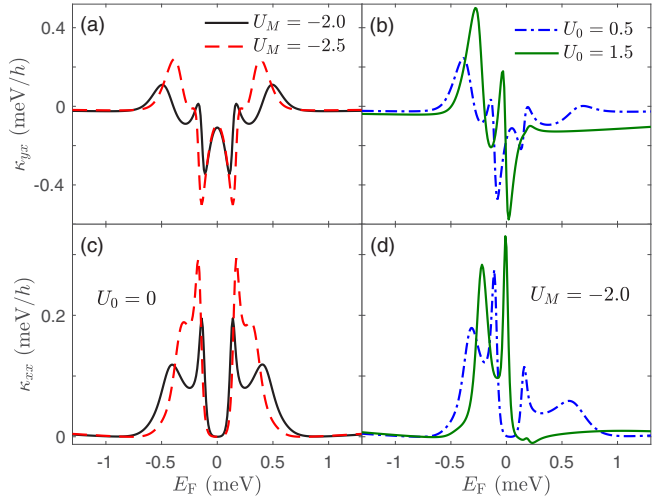

FIG. 5. The transverse (upper panel) and longitudinal (lower panel) thermal conductivity versus the Fermi level, with $U_0 = 0$ for the left panel and $U_M = -2.0$ for the right panel. Other parameters are the same as Fig. 4.

By substituting Eq. (9) into Eq. (34), we can derive $\mathbf{j}^Q = (\widehat{\overline{\kappa}} - \widehat{\overline{\alpha}}\widehat{\sigma}^{-1}\widehat{\alpha})(-\nabla T)$, so that the thermal conductivity tensor takes the form $\widehat{\kappa} = \widehat{\overline{\kappa}} - \widehat{\overline{\alpha}}\widehat{\sigma}^{-1}\widehat{\alpha}$. Consequently, we can further derive

$$\kappa_{xx} = \overline{\kappa}_{xx} - T\frac{\sigma_{xx}(\alpha_{xx}^2 - \alpha_{yx}^2) + 2\sigma_{yx}\alpha_{xx}\alpha_{yx}}{\sigma_{xx}^2 + \sigma_{yx}^2}, \quad (36)$$

$$\kappa_{yx} = \overline{\kappa}_{yx} - T\frac{\sigma_{yx}(\alpha_{yx}^2 - \alpha_{xx}^2) + 2\sigma_{xx}\alpha_{xx}\alpha_{yx}}{\sigma_{xx}^2 + \sigma_{yx}^2}. \quad (37)$$

At low temperatures, as can be seen from Eqs. (36) and (37), in the insulated regime, $\kappa_{xx} = \overline{\kappa}_{xx}$ and $\kappa_{yx} = \overline{\kappa}_{yx} - T\alpha_{yx}^2/\sigma_{yx}$, because $\sigma_{xx} = 0$. If $|E_F| \gg 0$, $\sigma_{yx} \to 0$, such that $\kappa_{xx} = \overline{\kappa}_{xx} - T\alpha_{xx}^2/\sigma_{xx}$ and $\kappa_{yx} = \overline{\kappa}_{yx}$. For the Fermi level between these two limits, both terms of Eqs. (36) and (37) contribute to the thermal conductivity, simultaneously. In Fig. 5, we plot the thermal conductivity tensor as functions of the Fermi level for different $U_M$ ($U_0$) in the left (right) panel. As can be seen, the thermal conductivity exhibits similar behaviors to the Nernst signal and thermopower, where both the longitudinal and transverse thermal conductivities are enhanced by the localization effect of the surface electrons, and suppressed when the Fermi level crosses the energy gap. The charge potential $U_0$, as shown by Figs. 5(b) and 5(d), can modify the resonant peaks of the thermal conductivity, in strength and energy positions, by renormalizing the localized levels.

In the above discussions, we assume $U_M < 0$ for the sake of brevity. The generalization of the conclusions to $U_M > 0$ is straightforward. With $U_M$ changing sign, the effective Zeeman-type field $\rho_{\text{imp}}\zeta_-(\epsilon^+)s_z$ changes its sign and, as a result, the anomalous Hall conductivity, the Nernst signal, and the transverse thermal conductivity will change sign as well.

In experiment, the energy locations of the localized states can be detected by the STM technology and the Nernst signal can be measured by a recent experiment setup developed to probe the spin-Seebeck effect (please see Ref. [53]). To detect the resonant enhancement of the ANS, one should ensure that the localized levels are within the energy range of the measurement. Therefore, before measuring the ANS, one should measure the DOS of the magnetically doped TI surface by STM, to identify if the localized levels emerge or not. The energy locations of the localized levels can be controlled by the doping density. After the localized levels have been observed in an appropriate energy, we can measure the transversal voltage $\Delta V_y$ upon application of a longitudinal temperature difference $\Delta T_x$. The ANS is determined by $S_{yx} = L_y^{-1}\Delta V_y/\Delta T_x$. By tuning the chemical potential, we can expect resonant peaks emerge in the ANS when the chemical potential passes through the localized levels. We should note that the resonant enhancement of the ANS is observable at relative low temperatures because thermal fluctuations can overwhelm the resonant peaks for higher temperatures.

## VI. SUMMARY

We generalized the Kubo-Streda formula to the calculation of the thermoelectric coefficients and studied the anomalous thermoelectric transport on a magnetically doped TI surface. It shows that the thermoelectric coefficients are simultaneously modulated by the impurity scattering and thermal fluctuations. The ANS and thermopower are enhanced around the magnetic-impurity-induced localized levels, while within the magnetic-impurity-induced gap, the ANS is fully suppressed and the thermopower is linearly scaled with the Fermi energy. The symmetrical distribution of the ANS and thermopower can be broken when a finite charge potential is included, making the signs of the ANS and thermopower tunable by the strength of the charge potential.

## ACKNOWLEDGMENT

This work was supported by the National Natural Science Foundation of China under Grants No. 11474106 (R.-Q.W), No. 11574155 (R.M.), No. 11804130 (W.L.), and the Key Program for Guangdong NSF of China under Grant No. 2017B030311003 (R.-Q.W) and GDUPS (2017).